\documentstyle[aps,epsf,prd]{revtex}
\begin{document}

\input amssym.def
\input amssym

\draft
\title{Nonassociative geometry: Towards discrete structure of spacetime }
\author{Alexander I. Nesterov}
\address{Departament of Physics, C.U.C.E.I., Guadalajara University,
Guadalajara, Jalisco, Mexico \\
and \\
L.V. Kirensky Institute of Physics, Siberian Branch Russian Academy of
Science, Krasnoyarsk, Russia \\
E-mail: nesterov@udgserv.cencar.udg.mx;
http://udgserv.cencar.udg.mx/$\sim$nesterov }

\author{ Lev V. Sabinin}
\address{Department of Mathematics, Quintana Roo University,
Chetumal, Mexico\\  and \\
Russian Friendship University, Moscow, Russia \\
E-mail: lsabinin@balam.cuc.uqroo.mx }

\wideabs{


\maketitle

\begin{abstract}
In the framework of {\em nonassociative geometry} a unified description of
continuum and discrete spacetime is proposed. In our approach at the Planck
scales the spacetime is described as a so-called {\em diodular discrete
structure} which at large spacetime scales ``looks like'' a differentiable
manifold. After a brief review of foundations of nonassociative geometry,
we discuss the nonassociative smooth and discrete de Sitter spacetimes.

\end{abstract}

}

Numerous attempts to construct the quantum theory of gravitation and
to understand the structure of spacetime has not been successful so far
and the problem is still open (for recent reviews see:
\cite{BI,I1,R1,R2,R3}). In general one should distinguish two  strategies
beyond the common treatment:
\begin{itemize}
\item Quantize a classical structure and then restore it as some kind of
the classical limit of the quantum theory.
\item Regard the classical structure as one being emerged from the other
theory.
\end{itemize}
The second strategy may require a revision of the quantum theory itself in
a way that that in quantum theory of gravitation the standard concept of
spacetime must be replaced at the Planck scales by some kind of the
discrete structure \cite{BI}.

In this paper we propose in the framework of nonassociative geometry
\cite{S5,S6,S7,S8,NS} a new approach to a classical and discrete structure
of spacetime, which provides the unified description of continuum and
discrete spacetime. The corresponding construction may be described as
follows. In a neighborhood of an arbitrary point on a manifold with an
affine connection one can introduce the geodesic local loop, which is
uniquely defined by means of the parallel translation of geodesics along
geodesics \cite{K,S1,S2}. The family of local loops constructed in this way
uniquely defines the space with affine connection, but not every family of
geodesic loops on a manifold defines an affine connection; there exist some
relations between the loops at distinct points. Taking into account that
the local loop sructure admits an additional operation, namely, the
multiplication of the point by scalar, and a vector space structure
induced by means of the exponential mapping; one can express the above
mentioned relations by means of some algebraic identities. This leads to
the notion of {\em odule} and the so-called {\em geoodular structure}.

A geoodular covering of the affinely connected manifolds, consisiting of
the odular covering with some additional algebraic identities, contains
complete information about the manifold and allows us to reconstruct it. If
we take an arbitrary smooth geoodular covering, then it uniquely generates
an affine connection, whose geoodular covering coincides with the initial
one. Introducing the left invariant diodular metric, one obtains the
Riemannian (pseudo-Riemannian) geodiodular manifold. This implies that a
smooth geoodular (Riemannian/pseudo-Riemannian) manifold is an affinely
connected (Riemannian/pseudo-Riemannian) manifold being described in
another language and there is the equivalence of the corresponding
categories \cite{S3,S4}. Ignoring the smoothness, it is possible to
consider discrete spaces, introduce a diodular ``metric'' and a
``connnection'' over arbitrary fields or rings, and even to define finite
spaces with ``affine connection''.

\subsection*{Nonassociative geometry in brief}

Here we survey algebraic foundations of {\em nonassociative
geometry} (for a recent review see: \cite{S8,NS}).\\

\noindent
{\bf Definition 1.} Let $\langle Q,{\mathbf\cdot}\rangle $ be a
groupoid with a binary operation $(a,b) \mapsto a {\mathbf\cdot} b$ and
$Q$ be a smooth manifold. Then $\langle Q,{\mathbf\cdot}\rangle $ is
called a {\it quasigroup} if each of the  equations $a{\mathbf\cdot}
x=b,~y{\mathbf\cdot} a=b$ has a unique solution:  $x=a\backslash b$,
$y=b/a$. A {\it loop} is a quasigroup with a two-sided identity,
$a{\mathbf\cdot} e= e{\mathbf\cdot} a=a, \forall a \in Q$. A loop
$\langle Q,{\mathbf\cdot},e \rangle$ with a smooth functions
$\phi(a,b):=a{\mathbf\cdot} b$ is called a {\it smooth loop}.

Let $\langle Q,{\mathbf\cdot},e\rangle$ be a smooth local loop with a neutral element $e$.
We define
\begin{equation}
L_a b=R_b a=a{\mathbf\cdot} b,\quad
l_{(a,b)}=L^{-1}_{a{\mathbf\cdot} b}\circ L_a\circ L_b,
\label{Ll}
\end{equation}
where $L_a$ is a {\it left translation}, $R_b$ is a {\it right
translation}, $l_{(a,b)}$ is an {\it associator}.\\

\noindent
{\bf Definition 2.} {Let $\langle{ M},{\mathbf\cdot},e \rangle$ be a
partial {\it groupoid} with a binary operation $(x,y)\mapsto
x{\mathbf\cdot} y$ and a neutral element $e ,\; x{\mathbf\cdot} e  =e
{\mathbf\cdot} x =x$; $   M$ be a smooth manifold (at least $C^1$-smooth)
and the operation of multiplication (at least $C^1$-smooth) be defined in
some neighborhood $U_e $,}. Then $\langle { M},{\mathbf\cdot},e \rangle$
is called a {\it partial loop on $   M$.} \\

\noindent
{\bf Definition 3.} {Let $\langle{ M},{\mathbf\cdot},e \rangle$ be a left
loop with a neutral element $e$ and $t: x \mapsto tx$ be a unary operation
such that $(t+u)x = tx{\mathbf\cdot} ux$, $(tu)x = t(ux)$ $(t,u \in
{\Bbb R}, \; x \in M)$. Then $\langle M,{\mathbf\cdot},e,(t)_{t\in
\Bbb R}\rangle $ is called a {\em left odule}.  } \\

\noindent
{\bf Definition 4.} {Let $M$} be a smooth manifold and
\[
L:  (x,a,y)\in {M\times M\times  M} \mapsto
x\;{}_{\stackrel{\mathbf\cdot}{a}}\;y= L^a_x y\in { M}
\]
a smooth partial ternary operation, such that $L^a_x y$ defines  in the
some neighbourhood of the point $a$   the loop with the neutral $a$. Then
the family $\bigl \{\langle  M,\;{}_{\stackrel{\mathbf\cdot}{a}}\;
\rangle\bigr\}_{a\in M}$ is called a {\it loopuscular structure}.

A smooth manifold $M$ with a smooth partial ternary operation $L$ and
smooth binary operations $t: (a,b)\in {M}\times {M} \mapsto
t_a b\in { M}, \;(t\in \Bbb  R)$, such that $L^a_x y$  and $t_a x$
determine in some neighborhood of an arbitrary point $a$ the odule with the
neutral element $a$, is called a {\it left odular structure}
$\bigl\{\langle  M,\;{}_{\stackrel{\mathbf\cdot}{a}}\;,(t_a)_{t\in
{\Bbb  R}} \rangle\bigr\}_{a\in M}$. If $\bigl\{\langle  M,\;
{}_{\stackrel{\mathbf\cdot}{a}}\;,(t_a)_{t\in {\Bbb  R}}
\;\rangle\bigr\}_{a\in M}$ and $\bigl\{\langle  M,\;
{}_{\stackrel{+}{a}}\;,(t_a)_{t\in {\Bbb  R}}
\;\rangle\bigr\}_{a\in M}$ are odular structures, then $\bigl\{\langle
M, \;{}_{\stackrel{\mathbf\cdot}{a}}\;,\;{}_{\stackrel{\mathbf
+}{a}}\;,(t_a)_{t\in {\Bbb R}} \;\rangle\bigr\}_{a\in M}$
is called a {\it diodular structure}. If $x\;{}_{\stackrel{\mathbf
+}{a}}\;y$ and $t_a x$
define a vector space, then such a diodular structure is called a {\it
linear diodular structure}.

A diodular structure is said to be {\it geodiodular} if
\begin{eqnarray}
&&\text{(the first geoodular identity)}\qquad
L_{u_ax}^{t_ax}\circ L_{t_ax}^{a}=L_{u_ax}^{a}, \nonumber\\
&&\text{(the second geoodular identity)} \quad
L_{x}^{a}\circ  {t_a}=t_x \circ  L_{x}^{a}, \nonumber\\
&&\text{(the third geoodular identity)} \quad \; \;
L_{x}^{a}(y\;{}_{\stackrel{\mathbf
+}{a}}\; z)= L_{x}^{a}y \;{}_{\stackrel{\mathbf
+}{a}}\;L_{x}^{a}z \nonumber
\end{eqnarray}
are true.

The equivalence of the categories of geoodular (geodiodular) structures
and of affine connections has been shown in \cite{S3,S4}. For a given
loopuscular structure the {\em tangent affine connection} is defined by
\begin{eqnarray}
&&\nabla_{X_a}Y=\left\{\frac{d}{dt}
\left([(L^a_{g(t)})_{*,a}]^{-1}Y_{g(t)}\right)\right\}_{t=0},\\
&&g(0)=a, \quad \dot g(0)=X_a,  \nonumber
\end{eqnarray}
$Y$ being a vector field in the neighborhood of a point $a$. In coordinates
the components of this affine connection are
\[
\Gamma^i_{jk}(a) = - \left[\frac{\partial^2 (L^a_x y)^i}{\partial
x^j \partial y^k}\right]_{x=y=a}.
\]
{\bf Definition 5.} {If $\bigl \{\langle  M,\;{}_{\stackrel{\mathbf\cdot}{a}}\;
\rangle\bigr\}_{a\in M}$ is a loopuscular structure then
\begin{equation}
h^a_{(b,c)} = (L^a_c)^{-1}\circ L^b_c \circ L^a_b
\label{hol}
\end{equation}
is called the {\em elementary holonomy}.}

The elementary holonomy is, in fact, some integral curvature. In the
smooth case, differentiating $(h^a_{(x,y)}z)^i$ by $x^l,\, y^k,\, z^j$ at
$a\in M$ we get the curvature tensor at $a\in M$  precisely up to a
numerical factor,
\begin{equation}
R^i{}_{jkl}(a)= 2 \left[\frac{\partial^3 (h^a_{(x,y)} z)^i}
{\partial x^l\partial y^k\partial z^j}\right]_{x=y=z=a}
\label{hl}
\end{equation}

{\bf Comment.} {Having the elementary holonomy $h_{(x,y)}$ at a neutral
element, one can obtain the diodular structure in a unique way \cite{S8}:
\begin{eqnarray}
&& L^a_x y = L_x^e h_{(a, x)}(L_a^e)^{-1}y, \\
&&x\; {}_{\stackrel{\mathbf +}{a}}\;y= L_a^e ( (L_a^e)^{-1}x\;
{}_{\stackrel{\mathbf +}{e}}\;(L_a^e)^{-1}y), \\
&& t_a y=L_a^e t_e(L_a^e)^{-1}y.
\end{eqnarray}

One may define left invariant metric $g^a(x,y)$ satisfying
\begin{equation}
g^b (L_b^a x,L_b^a y)=g^a(x,y).
\label{g0}
\end{equation}
In this way we obtain a {\it metric diodular structure} .

An odule $\langle M,\; {}_{\stackrel{\mathbf
+}{a}}\;, (t_a)_{t\in \Bbb
R}\rangle$  replaces the tangent space at a point $a$ and we call it an
{\em osculating space}. An element of the osculating space is called an
{\em osculating vector} and in a smooth case it is a geodesic arc starting
at the point $a$.  An {\em osculating structure}
$\bigl\{\langle  M,\; {}_{\stackrel{\mathbf
+}{a}}\;,(t_a)_{t\in {\Bbb  R}}
\;\rangle\bigr\}_{a\in M}$ plays the role of tangent bundle structure of
the smooth manifold. In the table below we compare the basic concepts of
the classical differential geometry and of the nonassociative geometry.

\vspace{0.1in}
\begin{center}
{\bf Differential Geometry Vs. Nonassociative Geometry}
\end{center}

\begin{center}
\begin{tabular}{|l|l|}\hline \hline
Differential Geometry & Non-associative Geometry \\ \hline \hline

Tangent space & Osculating space \\ \hline

Tangent bundle structure   & Osculating structure \\ \hline
Cotangent space & Co-osculating space \\ \hline

Parallel displacement & Left translations $L^a_x y$  \\ \hline

Curvature $R(X,Y)Z$& Elementary holonomy $h^a_{(x,y)}z$\\  \hline

Bianchi identities& Odular Bianchi identities\\ \hline \hline
\end{tabular}
\end{center}
\vspace{0.1in}

{\em Nonassociative discrete space.} Here we consider a geodiodular
discrete space ${\mathcal {M}}= \bigl\{\langle  M,\;
{}_{\stackrel{\mathbf\cdot }{a}}\;,{}_{\stackrel{\mathbf
+}{a}}\;,(t_a)_{t\in {\Bbb R}} \;\rangle\bigr\}_{a\in M}$.  This implies
that there exists geodiodular covering of $M$ and:

1. $\langle M,\;{}_{\stackrel{\mathbf \cdot}{a}}\;,(t_a)_{t\in \Bbb
R}\rangle $ is an odule with a neutral $a\in M$, for any $a\in M$,

2. $\langle M,\;{}_{\stackrel{\mathbf +}{a}}\;,(t_a)_{t\in \Bbb R}\rangle $
is a n-dimensional vector space (with zero element $a\in M)$,

3. The following geoodular identities
\begin{eqnarray}
&&\text{(the first geoodular identity)}\qquad
L_{u_ax}^{t_ax}\circ L_{t_ax}^{a}=L_{u_ax}^{a},\label{gid1}\nonumber \\
&&\text{(the second geoodular identity)} \quad
L_{x}^{a}\circ  {t_a}=t_x \circ  L_{x}^{a},\nonumber \\
&&\text{(the third geoodular identity)} \quad \; \;
L_{x}^{a}(y\;{}_{\stackrel{\mathbf +}{a}}\; z)= L_{x}^{a}y \;
{}_{\stackrel{\mathbf +}{x}}\; \;L_{x}^{a}z  \nonumber
\label{gid3}
\end{eqnarray}
are valid. One may consider the operations above as partially defined.

In our approach the osculating space ${\mathcal {M}}^+_a=\langle
M,{}_{\stackrel{\mathbf +}{a}}, a ,(t_a)_{t\in \Bbb R}\>$
plays the role of tangent space at $a\in M$. Indeed, in the smooth case the
tangent space $T_a M$ may be identified, at least locally, with $M$ by
means of the exponential map. Any point $x\in M$ may be, then, regarded as
an osculating vector in ${\mathcal {M}}_a^+ \; (\forall a \in M)$. Any line
$(t_a b)_{t\in \Bbb R}$ may be regarded as a geodesic through $a,b \in
M$.

The presence of curvature in such a space results in a non-trivial
elementary holonomy,
\[
L_a^y\circ  L_y^x\circ  L_x^a= h^a(x,y)\neq \rm {Id}.
\]

Having given a discrete nonassociative space (finite affinely connected
space in other words) we can enrich it by a {\em metric diodular
structure}. Namely, one may define additionally non-degenerate left
invarint metric $g^a (x,y)$ for any ${\mathcal{ M}}_{a}^+$:
$ g^b (L_b^a x,L_b^a y)=g^a(x,y)$.

\subsection*{Nonassociative de Sitter spacetime}

The spacetimes of constant curvature are locally characterized by the
condition
\begin{equation}
R_{\mu\nu\lambda\sigma}= K(g_{\mu\lambda}g_{\nu\sigma} -
g_{\mu\sigma}g_{\nu\lambda}).
\label{R}
\end{equation}
One can regard these spaces as solutions of the Einstein's equations with
$\Lambda$-term for an empty space and $\Lambda = 3K$.

The spacetime of constant curvature with $K > 0$ is called {\em de Sitter
spacetime}, and if $K<0$ it is called {\em anti-de Sitter spacetime}.
De Sitter spacetime has the topology $S^3\times {\Bbb R}$ and can be
considered as hyperboloid \cite{HE,KR}
\[
Z_a Z^a -Z_0Z^0=-1/K, \; a =
1,\dots,4
\]
embedded into five-dimensional Minkowski space with metric \[
ds^2 = (dZ^0)^2 - dZ_adZ^a.
\]
In the remainder of this section we consider de Sitter spacetime with the
inner metric \cite{KR}
\begin{equation}
ds^2  = \frac{dt^2 - dx^2-dy^2-dz^2}{\left(1-
\frac{K}{4}(t^2-x^2-y^2-z^2)\right)^2},
\label{g1}
\end{equation}
but our approach may be extended to anti-de Sitter spacetime as well.

\vspace{0.1in}
{\em Smooth de Sitter spacetime.} The easiest way to study the spacetimes
of the constant curvature in the framework of nonassociative geometry is
to employ the quaternions. Let us consider the quaternionic algebra over
complex field ${\Bbb C}(1,\rm i)$
\[
{\sf H}_{\cal C} = \{ \alpha + \beta i+ \gamma j + \delta k\; \mid
 \; \alpha , \beta,\gamma,\delta \in {\Bbb C} \}
\]
with multiplication operation defined by the property of bilinearity and
following rules for $i, \; j, \; k$:
\begin{eqnarray}
&&i^{2}=j^{2}=k^{2} =-1,\;\; jk=-kj = i, \nonumber\\
&&ki=-ik=j,\;\; ij=-ji= k.\nonumber
\end{eqnarray}
The quaternionic conjugation (denoted by $^{+}$) is defined by
\begin{equation}
q^+=\alpha - \beta i -\gamma j - \delta k
\end{equation}
for $q=\alpha + \beta i+ \gamma j + \delta k$.
This definition implies
\begin{equation}
(q p)^+ = p^+ q^+, \quad p,q\in{\sf H_{\cal C}}.
\end{equation}

Further we restrict ourselves to the set of quaternions
 ${\sf H}_{\Bbb R}$:
\begin{eqnarray}
{\sf H}_{\Bbb R} =&&\{\zeta=\zeta^0 + {\rm i}(\zeta^1 i+ \zeta^2 j+ \zeta^3
k):\; \;{\rm i}^2=-1,\;\rm i\in{\Bbb C}, \nonumber \\
&&\zeta^0,\zeta^1,\zeta^2,\zeta^3  \in {\Bbb R} \}  \nonumber
\end{eqnarray}
 with the norm $\|\zeta\|^2$ given by
\begin{equation}
\|\zeta\|^2=\zeta {\zeta}^+=(\zeta^0)^2 - (\zeta^1)^2-(\zeta^3)^2 -
 (\zeta^4)^2.
\end{equation}

Introducing a binary operation
\begin{equation}
\zeta\ast\eta=\big(\zeta+\eta\big)\Big/\Big(1 + \frac{K}{4}\zeta^+\eta
\Big), \; \zeta,\eta \in {\sf H}_{\Bbb R} \label{QL}
\end{equation}
where $K$ is constant and $/$ denotes the right division, we find that the set of
quaternions ${\sf H}_{\Bbb R}$ with the binary operation $\ast$ forms a
loop $Q{\sf H}_{\Bbb R}$ which admits a natural geodiodular structure
induced by the quaternionic algebra. The associator is found to be
\begin{equation}
l_{(\zeta,\eta)}\xi= \Big(1+\frac{K}{4}\zeta\eta^+ \Big)\xi
\Big/\Big(1+\frac{K}{4}\zeta^+\eta \Big).
\end{equation}

Employing (\ref{g0}), we define the left invariant diodular metric on
$Q{\sf H}_{\Bbb R}$ as follows:
\begin{eqnarray}
g^{\zeta}(\xi,\eta) = \frac{(\xi-\zeta)(1 - \frac{K}{4}\xi^{+}\zeta )(1 -
\frac{K}{4}\zeta^{+}\eta )(\eta^{+ } - \zeta^{+})}{2\|1 -
\frac{K}{4}\zeta^{+}\xi \|^2 \;\|1- \frac{K}{4}\zeta^{+}\eta \|^2}  +
\nonumber\\
+\frac{(\eta-\zeta)(1 - \frac{K}{4}\eta^{+}\zeta )(1 -
\frac{K}{4}\zeta^{+}\xi)(\xi^{+ } - \zeta^{+})} {2\|1 -
\frac{K}{4}\zeta^{+}\xi \|^2 \;\|1- \frac{K}{4}\zeta^{+}\eta
\|^2}.\nonumber \label{dg}
\end{eqnarray}
In particular,
\begin{equation}
g^{0}(\xi,\eta) =\frac{1}{2}(\xi\eta^{+} + \eta\xi^{+})=\xi^\nu\eta_\nu,
\end{equation}
and
\begin{equation}
g^{\zeta}(\xi,\xi) = \frac{\|\xi-\zeta\|^2}{\|1 -\frac{K}{4}\zeta^{+}\xi \|^2 } .
\label{dm}
\end{equation}
Let $\xi = \zeta + d \zeta $, then (\ref{dm}) leads to the de Sitter metric
(\ref{g1}):
\[
g(d\zeta,d\zeta) =
\frac{(d\zeta^0)^2-(d\zeta^1)^2-(d\zeta^2)^2-(d\zeta^3)^2} {\bigl(1 -
\frac{K}{4}((\zeta^0)^2-(\zeta^1)^2-(\zeta^2)^2-(\zeta^3)^2)\bigr)^2}.
\]

Taking into account that for symmetric spaces its elementary holonomy is
determined by the associator \cite{S5,S8}:
$$
h_{(\zeta,\eta)} \xi= l_{(\zeta,L^{-1}_\zeta \eta)}\xi,
$$
we find
\begin{equation}
h_{(\zeta,\eta)}\xi = \Big(1- \frac{K}{4}\eta \zeta^{+} \Big )\xi
\Big /\Big(1- \frac{K}{4}\eta^{+}\zeta \Big).
\label{hol01}
\end{equation}

Applying (\ref{hl}), we obtain the curvature tensor of de Sitter spacetime
in the normal coordinates as the following:
$$
R_{\mu\nu\lambda\sigma}=- \frac{K}{2}\varepsilon_{\mu\nu\kappa\delta}
\varepsilon^{\kappa\delta} {}_{\lambda\sigma}.
$$
\noindent
{\bf Comment.} The de Sitter spacetime may be obtained as a
solution of the diodular Einstein's equations with `$\Lambda$-term'
\cite{NS,NS1}.

\vspace{0.1in}
{\em Discrete de Sitter spacetime.} Let us consider a finite set $M = {\Bbb
Z}^4_n = \{ {\bf p} =(p^\mu)\;|\; p^\mu \in {\Bbb Z}_n, \mu = 0,\dots,3,
n\in{\Bbb N}\}$ where ${\Bbb Z}_n = \{p=-n,\dots,n \}$ is the set of
integers. We define a partial loop  $Q{\sf H}_{{\Bbb Z}^4_n}$ as a set of
quaternions
\begin{eqnarray}
{\sf H}_{{\Bbb Z}^4_n} =&&\{\zeta_{\bf p}=\ell(p^0 + {\rm i}
(p^1 i+ p^2 j+ p^3 k)): \; \ell = {\rm const}, \nonumber \\
&&{\rm i}^2=-1,\;\rm i\in{\Bbb C};\;
{\bf p}  \in {\Bbb Z}^4_n\}
\end{eqnarray}
with the indefinite norm $\|\zeta_{\bf p}\|^2=\ell^2 p^\mu p_\mu$ and the
binary operation ${\sf H}_{{\Bbb Z}^4_n}\times {\sf H}_{{\Bbb Z}^4_n}
\mapsto {\sf H}_{{\Bbb Z}^4_n} $ defined by
\begin{equation}
\zeta_{\bf p}\ast\zeta_{\bf q}=\zeta_{\bf pq} =\Big ({\zeta_{\bf
p}+\zeta_{\bf q}}\Big )\Big/\Big({1+ \frac{K}{4}\zeta^{+}_{\bf
p}\zeta_{\bf q} }\Big),\; \; \zeta_{\bf p},\zeta_{\bf q} \in {\sf
H}_{{\Bbb Z}^4_n} \label{DLp}.  \end{equation}

We introduce a partial geodiodular finite space ${\cal M}_n $ at the
neutral element $e$ (zero) as follows:
\begin{itemize}
\item $Q{\sf H}_{{\Bbb Z}^4_n}$ being the odule with the unary operation
multiplication induced by the quaternionic algebra over $\Bbb Z$

\item ${\cal M}^{+}= {\sf H}_{{\Bbb Z}^4_n}$ being the osculating space
with the structure of vector space induced by the quaternionic algebra over
$\Bbb Z$.
\end{itemize}
Employing the geoodular identities one can obtain
the geoodular covering of the discrete spacetime.

\noindent

For the duiodular metric and elementary holonomy we have
\begin{eqnarray}
& g^{\zeta_{\bf p}}(\zeta_{\bf q},\zeta_{\bf q}) =
\frac{\|\zeta_{\bf p}-\zeta_{\bf q}\|^2}
{\|1 - \frac{K}{4}\zeta^{+}_{\bf p}\zeta_{\bf q} \|^2 },\label{dm2} \\
& h_{(\zeta_{\bf p},\zeta_{\bf q})}\zeta_{\bf m}
=\Big(1- \frac{K}{4}\zeta_{\bf q}\zeta^{+}_{\bf p} \Big )
\zeta_{\bf m}\Big/ \Big(1- \frac{K}{4}\zeta^{+}_{\bf q}\zeta_{\bf p} \Big ).
\label{hol2}
\end{eqnarray}

The smooth spacetime could be regarded as the result of ``limit process
of triangulating'' when  $\ell n = \rm const$, while $\ell \longrightarrow
0 ,\; n \longrightarrow \infty$. Let us consider ${\bf q = p +
\mbox{\boldmath $\delta$}}, |\mbox{\boldmath $\delta$} | \ll n$. Then we
have $\zeta_{\bf q} = \zeta_{\bf p} + \Delta \zeta_{\bf q}$ where $\Delta
\zeta_{\bf q} = \ell(\delta^0 + {\rm i}(\delta^1 i+ \delta^2 j+ \delta ^3
k)) $, and the diodular metric (\ref{dm2}) takes the form
\begin{equation}
g^{\zeta_{\bf p}}(\zeta_{\bf q},\zeta_{\bf q}) =
\frac{\|\Delta\zeta_{\bf q}\|^2}
{\|1 - \frac{K}{4}\zeta^{+}_{\bf p}\zeta_{\bf p} \|^2 }+ O(K\ell^2),
\label{dm4}
\end{equation}
and
\begin{eqnarray}
g^{\zeta_{\bf p}}(\zeta_{\bf
q},\zeta_{\bf q}) \longrightarrow
ds^2  = \frac{(d\zeta^0)^2-(d\zeta^1)^2-(d\zeta^2)^2-(d\zeta^3)^2}
{(1 - \frac{K}{4}((\zeta^0)^2-(\zeta^1)^2-(\zeta^2)^2-(\zeta^3)^2)^2},
\nonumber
\end{eqnarray}
while $\ell \rightarrow 0,\; n \rightarrow 0$.
Comparing  with the smooth case, we see that to some extent the information
concerning the geometry of de Sitter spacetime is hidden in the
structure of finite loop.

The similar consideration of the elementary holonomy gives
\begin{equation}
h_{(\zeta_{\bf p},\zeta_{\bf q})}\zeta_{\bf m}
=\zeta_{\bf m} + \frac{K}{4}\Delta(\zeta_{\bf p},\zeta_{\bf q},\zeta_{\bf
m}) + O \left(K\ell^2\right) \label{hol3},
\end{equation}
where
\[
\Delta(\zeta_{\bf p},\zeta_{\bf q},\zeta_{\bf m})
= \zeta_{\bf m}\zeta^{+}_{\bf q}\zeta_{\bf p}
- \zeta_{\bf q}\zeta^{+}_{\bf p}\zeta_{\bf m}.
\]
In the coordinates (\ref{hol3}) can be written as
\[
\left(h_{(\zeta_{\bf p},\zeta_{\bf q})}\right)^\mu_\nu\zeta^\nu_{\bf m}
=\zeta^\mu_{\bf m} + \frac{K}{4}
\varepsilon_{\mu\nu\kappa\delta}
\varepsilon^{\kappa\delta} {}_{\lambda\sigma}
\zeta^\nu_{\bf m}\zeta^\lambda_{\bf p}\zeta^\sigma_{\bf q}
+ O \left(K\ell^2\right)
\]
Approaching the limit, while
$n\longrightarrow \infty, \ell \longrightarrow 0$, one restores
the curvature tensor of de Sitter spacetime in the normal coordinates:
$$
R_{\mu\nu\lambda\sigma}= -\frac{K}{2}\varepsilon_{\mu\nu\kappa\delta}
\varepsilon^{\kappa\delta} {}_{\lambda\sigma}.
$$
~~~~~~\\

The above examples show how the continuum and discrete structure of
spacetime might be described in the framework of the nonassociative
geometry. We deal with points and essentially nonassociative operation.
This leads us to the concept of the nonassociative (discrete) spacetime,
when at distances comparables with Planck length the standard concept of
spacetime might be replaced by the diodular discrete structure which at
large spacetime scales ``looks like'' a differentiable manifold.


\begin{thebibliography}{99}

\bibitem{BI} J. Butterfield and C.J.Isham, {\em Spacetime and the
Philosophical Challenge of Quantum Gravity,} gr-qc/9903072.

\bibitem {I1} C.J. Isham, {\em Structural Issues in Quantum Gravity,}
gr-qc/9510063.

\bibitem {R1} C. Rovelli, {\em Quantum spacetime: what do we know?}
gr-qc/9903045.

\bibitem {R2} C. Rovelli, {\em Loop Quantum Gravity.}Living Reviews in
Relativity (refereed electronic journal),
http://www.livingreviews.org/Articles/Volume1/1998-1rovelli; gr-qc/9709008.

\bibitem {R3} C. Rovelli, {\em String, loops and others: a critical
survey of the presence approaches to quantum gravity,} gr-qc/9803024.

\bibitem {S5} L.V. Sabinin, {\it Differential equations of smooth
loops}, in:  {Proc. of Sem. on Vector and Tensor Analysis,} {\bf 23} 133.
Moscow:  Moscow Univ. 1988.

\bibitem {S6} L.V. Sabinin, {\it  Differential Geometry and Quasigroups},
Proc. Inst. Math. Siberian Branch of Ac. Sci. USSR, {\bf 14}, 208 (1989).

\bibitem {S7} L.V. Sabinin, {\it On differential equations of smooth
loops}, Russian Mathematical Survey, {\bf 49} 172 (1994).

\bibitem {S8} L.V. Sabinin, {\it Smooth quasigroups and loops,}
Dordrecht: Kluwer Academic Publishers (1999).

\bibitem {NS} A.I. Nesterov and L.V. Sabinin, {\em Non-aasociative geometry
and discrete structure of spacetime,} Comment. Math. Univ.  Carolinae {\bf
41,2}, 347 (2000), hep-th/0003238.

\bibitem {NS1} A.I. Nesterov, and L.V. Sabinin, {in preparation}.

\bibitem {K} M. Kikkawa, {\it On local loops in affine manifolds}, J.
Sci. Hiroshima Univ. Ser A-I Math. {\bf 28}, 199 (1961).

\bibitem {S1} L.V. Sabinin, {\it  The geometry of loops}, Mathematical
Notes, {\bf 12}, 799 (1972).

\bibitem {S2} L.V. Sabinin, {\it  On the equivalence of categories of
loops and homogeneous spaces}, Soviet Math. Dokl. {\bf 13}, 970 (1972).

\bibitem {S3} L.V. Sabinin, {\it  Odules as a new approach to a
geometry with a connection,} Soviet Math. Dokl. {\bf 18}, 515 (1977).

\bibitem {S4} L.V. Sabinin, {\it Methods of Nonassociative Algebra in
Differential Geometry}, in {Supplement to Russian translation of S.K.
Kobayashi and K. Nomizy ``Foundations of Differential Geometry'', Vol. 1.}
Moscow: Nauka (1981).

\bibitem {HE} S.W. Hawking and G.F.R. Ellis, {\em The large scale
structure of space-time.} Cambridge: Univ. Press. (1975)

\bibitem {KR} D. Kramer, H. Stephani, M. MacCallum and E. Herlt,{\em Exact
Solutions of Einstein's Equations,} Berlin: VEB Deutscher Verlag (1980).

\end{thebibliography}
\end{document}